\documentclass[12pt]{article}
\usepackage{amsmath,amssymb}
\usepackage{graphicx,color}
\numberwithin{equation}{section}
\usepackage{cite}
\usepackage{bm}
\usepackage{dcolumn}  
\newcommand{\be}{\begin{equation}}
\newcommand{\bea}{\begin{eqnarray}}
\newcommand{\eea}{\end{eqnarray}}
\newcommand{\ba}{\begin{array}}
\newcommand{\ea}{\end{array}}
\newcommand{\ee}{\end{equation}}

\expandafter\ifx\csname mathbbm\endcsname\relax

\else

\fi \textheight 22cm \textwidth 15cm \topmargin 1mm
\def\N{{\cal \nu}}
\begin{document}
\begin{titlepage}
\hfill
\vbox{
    \halign{#\hfil         \cr
           IPM/P-2007/011 \cr
                      } 
      }  
\vspace*{20mm}
\begin{center}
{\Large {\bf On $R^2$ Corrections for 5D Black Holes }\\
}

\vspace*{15mm}
\vspace*{1mm}
{Mohsen Alishahiha\footnote{alishah@theory.ipm.ac.ir}}
\vspace*{1cm}

{\it  Institute for Studies in Theoretical Physics
and Mathematics (IPM)\\
P.O. Box 19395-5531, Tehran, Iran \\ \vspace{3mm}}

\vspace*{1cm}
\end{center}

\begin{abstract}
We study higher order corrections to extremal black holes/black string in five dimensions. 
These higher order corrections are due to supersymmetric completion of $R^2$ term in 
five dimensions. By making use of the results we extend the notion
of very special geometry when higher derivative terms are also taken into account. This can
be used to make a connection between total bundle space of near horizon wrapped M2's and 
wrapped M5's in the presence  of higher order corrections. We also show how the corrected 
geometry removes the singularity of a small black hole. 
\end{abstract}
\end{titlepage}

\section{Introduction}

In recent years our understanding of corrections to the black hole entropy has increased considerably.
In a gravitational theory  using the Wald entropy formula \cite{Wald:1993nt} one can find the 
contribution of higher order corrections to the tree level Bekenstein-Hawking area law formula.
For extremal black holes in string theory taking into account the higher derivative terms
the corrected entropy has been evaluated in several papers including \cite{{Behrndt:1998eq},
{Dabholkar:2004yr},{Sen:2004dp}}
where it was shown that the results are in agreement with microscopic description of entropy
coming from microstate counting in string theory.

Although at the beginning the corrections have mainly been studied  for the supersymmetric black holes 
where the attractor mechanism \cite{Ferrara:1995ih}
was observed, recently it was generalized to non-supersymmetric cases as well
(for example see \cite{{Sen:2005iz},{Goldstein:2005hq},{Tripathy:2005qp},{Prester:2005qs},
{Kallosh:2006bt},{Alishahiha:2006ke},{Chandrasekhar:2006kx},{Exirifard:2006wa},
{Ghodsi:2006cd},{Cai:2006xm},{Astefanesei:2006dd},{Kaura:2006mv},{Morales:2006gm},
{Arfaei:2006qr},{Astefanesei:2006sy},{Chandrasekhar:2006ic},{Andrianopoli:2006ub},{Cardoso:2006xz}}).

An extremal black hole in four dimensions  has $AdS_2\times S^2$ near horizon geometry, while in five 
dimensions the near horizon geometry could be either $AdS_2\times S^3$ or $AdS_3\times S^2$. To compute
the contributions of higher order corrections to a black hole with near horizon geometry $AdS_2\times S^3$ 
one may 
simplify the Wald formula leading to the entropy function formalism \cite{Sen:2005wa}. 
While for those with $AdS_3\times S^2$ near horizon geometry it is useful to work within
the c-extremization framework \cite{Kraus:2005vz}. Using these mechanisms one can find corrections
to the Bekenstein-Hawking area law formula coming from higher derivative terms. These corrections
have to be compare with the microstate counting in string theory/M-theory. 

As far as the five dimensional black holes are concerned, We note that although the microscopic 
origin of ${\cal N}=8,4$ five dimensional rotating black
holes has been understood for a decade \cite{{Strominger:1996sh},{Breckenridge:1996is}}, the origin of 
the entropy for five dimensional ${\cal N}=2$ black hole has not been fully understood yet. Recently
the microscopic accounting of the five dimensional rotating black hole arising from wrapped M2-branes in Calabi-Yau
compactification of M-theory has been studied in \cite{Guica:2007gm} where the authors established a connection
between this black hole and a well understood one by making use of an embedding of space-time in the 
total space of the $U(1)$ gauge bundle over near horizon geometry of the black holes. 
Of course it was done in an specific case, namely near zero-entropy, zero-temperature and maximally 
rotating limits. 

This is the aim of this article to further study higher derivative corrections to ${\cal N}=2$ five dimensional
black holes. To do this we will work with the full 5D supersymmetry invariant four-derivative action, corresponding to the supersymmetric completion of the four-derivative Chern-Simons term which has recently been 
obtained in \cite{Hanaki:2006pj}.

The article is organized as follows. In section 2 we shall fix our notation where we present the
five dimensional action obtained in \cite{Hanaki:2006pj}. In section 3 we will consider a five
dimensional black string at $R^2$ level. In section 4 we will generalize the notion of very
special geometry in the presence of higher derivative terms using the results of section 3. 
This generalization can be used to extend the consideration of \cite{Guica:2007gm} to the case 
where higher order corrections are also taken into account. 
In section 5 we shall study higher order corrections to a
five dimensional extremal black hole using the fully supersymmetrized higher derivative terms. 
We then compare the result with the corrections coming from the bosonic Gauss-Bonnet action. 
We shall also see how the higher derivative terms remove
the singularity of a small black hole.  The last section is devoted to  discussions.

\section{Basic setup}

In this section we will fix our notation. We would like to study ${\cal N}=2$ supergravity in five
dimensions. This model is usually studied in the context of very special geometry. We note, however, that
sometime it is useful to work with a more general context, namely the superconformal approach.
This approach, in particular, is useful when we want to write
the explicit form of the action.  
In this approach we start with a five dimensional theory which is invariant under a
larger group that is superconformal group and therefore we construct a conformal supergravity.
Then by imposing a gauge fixing condition, one breaks the conformal supergravity to 
standard supergravity model. 

The representation of superconformal group includes Weyl, vector and
hyper multiples. The 
bosonic part of the Weyl multiplet contains the vielbein $e_\mu^a$, two-form auxiliary
field $v_{ab}$, and a scalar auxiliary field $D$. The bosonic part of the vector multiplet 
contains one-form gauge field $A^I$ and scalar fields $X^I$, where $I=1,\cdots,n_v$ labels
generators of a gauge group. The hypermultiplet contains scalar fields ${\cal A}_\alpha^i$
where $i=1,2$  is $SU(2)$ doublet index and $\alpha=1,\cdots, 2r$ refers to $USp(2r)$ group.
Although we won't couple the theory to matters, we shall consider the hyper multiplet to gauge fix 
the dilataional symmetry reducing the action to the standard ${\cal N}=2$ supergravity action. 

In this notation at leading order the bosonic part of the action is\cite{Hanaki:2006pj}
\be
I=\frac{1}{16\pi G_5}\int d^5x{\cal L}_0,
\ee
with
\bea
{\cal L}_0&=&\partial_a{\cal A}_\alpha^i\partial^a{\cal A}^\alpha_i+
(2\N+{\cal A}^2)\frac{D}{4}+(2\N-3{\cal A}^2)\frac{R}{8}+(6\N-{\cal A}^2)\frac{v^2}{2}
+2\N_IF_{ab}^Iv^{ab}
\cr &+&\frac{1}{4}\N_{IJ}(F^I_{ab}F^{J\;ab}+2\partial_a X^I\partial^a X^J)
+\frac{g^{-1}}{24}C_{IJK}\epsilon^{abcde}A^I_a F^J_{bc}F^K_{de},
\label{5acts}
\eea
where ${\cal A}^2={\cal A}_{\alpha\;ab}^i{\cal A}^{\alpha\;ab}_i,\;v^2=v_{ab}v^{ab}$ and
\be
\N=\frac{1}{6}C_{IJK}X^IX^JX^K,\;\;\;\;\;\;\;\N_I=\frac{1}{2}C_{IJK}X^JX^K,\;\;\;\;\;\;\;
\N_{IJ}=C_{IJK}X^K.
\ee

To fix the gauge it is convenient to set ${\cal A}^2=-2$. Then integrating out the auxiliary fields
by making use of their equations of motion one finds
\be
{\cal L}_0=R -\frac{1}{2}G_{IJ}F^I_{ab}F^{Jab}-
{\cal G}_{ij} \partial_a \phi^i\partial^a \phi^j+\frac{g^{-1}}{24}\epsilon^{abcde}C_{IJK}
F_{ab}^IF_{cd}^JA^K_e.
\label{5act}
\ee
The parameters in the action (\ref{5act}) are defined by
\be
G_{IJ}=-\frac{1}{2}\partial_I\partial_J\log {\cal \nu}|_{{\cal \nu}=1}\;,\;\;\;\;\;\;\;\;\;\;\;\;
{\cal G}_{ij}=G_{IJ}\;\partial_iX^I\partial_jX^J|_{{\cal \nu}=1}\;,
\ee
where $\partial_i$ refers to a partial derivative with respect to the scalar fields $\phi^i$.
In fact doing this, we recover the very special geometry underlines the theory in the leading order.

We can also find higher derivative terms in the action using the superconformal language. 
Actually, the supersymmetrized higher order action with four-derivative has recently been obtained 
in \cite{Hanaki:2006pj}. The corresponding action
is   
\bea\label{act5dR2}
{\cal L}_1&=&\frac{c_{2I}}{24}\bigg(\frac{1}{16}g^{-1}\epsilon_{abcde}A^{Ia}C^{bcfg}
C^{de}_{\phantom{de}fg}+\frac{1}{8}X^IC^{abcd}C_{abcd}+\frac{1}{12}X^ID^2
+\frac{1}{6}F^{Iab}v_{ab}D\cr 
&&\;\;\;\;\;\;\;\;-\frac{1}{3}X^IC_{abcd}v^{ab}v^{cd}
-\frac{1}{2}F^{Iab}C_{abcd}v^{cd}+\frac{8}{3}X^Iv_{ab}\hat{{\cal
D}}^b\hat{{\cal D}}_cv^{ac}\\ \nonumber
&&\;\;\;\;\;\;\;\; +\frac{4}{3}X^I\hat{{\cal
D}}^av^{bc}\hat{{\cal D}}_av_{bc}+\frac{4}{3}X^I\hat{{\cal
D}}^av^{bc}\hat{{\cal
D}}_bv_{ca}-\frac{2}{3}e^{-1}X^I\epsilon_{abcde}v^{ab}v^{cd}\hat{{\cal
D}}_fv^{ef}\\ \nonumber
&&\;\;\;\;\;\;\;\;+\frac{2}{3}e^{-1}F^{Iab}\epsilon_{abcde}v^{cd}\hat{{\cal
D}}_fv^{ef}+e^{-1}F^{Iab}\epsilon_{abcde}v^c_{\phantom{c}f}\hat{{\cal
D}}^dv^{ef} \cr
&&\;\;\;\;\;\;\;\; -\frac{4}{3}F^{Iab}v_{ac}v^{cd}v_{db}
-\frac{1}{3}F^{Iab}v_{ab}v^2+4X^Iv_{ab}v^{bc}v_{cd}v^{da}-
X^I(v_{ab}v^{ab})^2\bigg),\nonumber
\eea
where $C_{abcd}$ is the Weyl tensor defined as
\be
C^{ab}_{\phantom{ab}cd}=R^{ab}_{\phantom{ab}cd}+\frac{1}{6}R\delta^{[a}_{\phantom{a}[c}\delta^{b]}_{\phantom{b}d]}
-\frac{4}{3}\delta^{[a}_{\phantom{a}[c}R^{b]}_{\phantom{b}d]}~.
\ee
The double covariant derivative of $v_{ab}$ has curvature contributions given by
\be
v_{ab}\hat{{\cal D}}^b\hat{{\cal D}}_cv^{ac}=v_{ab}{\cal
D}^b{\cal
D}_cv^{ac}+\frac{2}{3}v^{ac}v_{cb}R_{a}^{\phantom{a}b}+\frac{1}{12}v_{ab}v^{ab}R~.
\ee

In general it is quite difficult to solve the equations of motion coming from the action 
contains both ${\cal L}_0$ and ${\cal L}_1$. Nevertheless as long as the supersymmetric solutions
are concerned, it is useful to look at the supersymmetry transformations which could give a simple way
to find explicit solutions. The supersymmetry variations of the fermions in Weyl, vector and hyper
multiplets are
\bea
\delta \psi_\mu^i&=&{\cal D}_\mu\varepsilon^i+\frac{1}{2}v^{ab}\gamma_{\mu ab}\varepsilon^i-\gamma_\mu\eta^i,\cr
\delta \chi^i&=&D\varepsilon^i-2\gamma^c\gamma^{ab}\varepsilon^i{\cal D}_av_{bc}-2\gamma^a\varepsilon^i
\epsilon_{abcde}v^{bc}v^{de}+4\gamma\cdot v\eta^i,\cr
\delta\Omega^{Ii}&=&-\frac{1}{4}\gamma\cdot F^I\varepsilon^i-\frac{1}{2}\gamma^a\partial_aX^I-X^I\eta^i,\cr
\delta\zeta^\alpha&=&\gamma^a\partial_a{\cal A}_j^\alpha-\gamma\cdot v\varepsilon{\cal A}^\alpha_j+3{\cal A}_j^\alpha
\eta^i,
\label{SUSY}
\eea
where garavitino $\psi^i_\mu$ and the auxiliary Majorana spinor $\chi^i$ come from the Weyl multiple, while
the gaugino $\Omega^{Ii}$ and $\zeta^\alpha$ come from  vector and huper multiplets, respectively.

\section{Black string and c-extremization }\label{3}

In this section we will consider a five dimensional extremal black string solution 
whose near horizon geometry is $AdS_3\times S^2$. Using the symmetry of the near horizon geometry one may 
start with the following ansatz for near horizon solution
\be
ds=l_A^2ds_{ADS_3}^2+l_S^2ds_{S^2}^2,\;\;\;\;\;X^I=cont.\;\;\;\;\;\;F^I_{\theta\phi}=\frac{p^I}{2}\sin\theta.
\label{ads3s2}
\ee
By making use of the c-extremization \cite{Kraus:2005vz} we can fix the parameters $l_A,l_S$ and $X_I$ as follows.
In this method we will first define c-function whose critical points correspond to the solutions of the
equations of motion. Then evaluating the c-function at critical points gives the average of the
left and right moving central charges of the associated CFT. 

In our model at leading order the c-function 
is given by 
\be
c=-6\;l_A^3l_S^2\; {\cal L},
\label{C}
\ee
where ${\cal L}$ is the Lagrangian evaluated on the above ansatz. The parameters of the 
ansatz are obtained by extremizing this function with respect to them
\be
\frac{\partial c}{\partial l_A}=0,\;\;\;\;\;\;\;\;\;\;\frac{\partial c}{\partial l_S}=0,
\;\;\;\;\;\;\;\;\;\;\frac{\partial c}{\partial X^I}=0.
\ee
For the five dimensional action (\ref{5act}) the above equations can be solved 
leading to
\be
l_A=2l_S=(\frac{1}{6}C_{IJK}p^Ip^Jp^K)^{1/3},\;\;\;\;\;\;\;\;\;\;X^{I}=\frac{p^I}{(\frac{1}{6}C_{IJK}p^Ip^Jp^K)^{1/3}}.
\ee 
Plugging these into (\ref{C}), one finds the central charge for the black string
at two-derivative level as $c=C_{IJK}p^Ip^Jp^K$. 

Let us now redo c-extremization for the black string solution (\ref{ads3s2}) in the presence of higher derivative
terms. This correction has recently been studied in \cite{CD5}.
In general one should evaluate the Lagrangian (\ref{5acts}) and (\ref{act5dR2}) for our ansatz
(\ref{ads3s2}) and then extremize it with respect to the parameters. We note, however, that it is in general
difficult to solve the equations of motion explicitly. Nevertheless one may use the supersymmetry
transformations (\ref{SUSY}) to fix some of the parameters. The remaining parameters can then be found
by equation of motion of the auxiliary  field $D$. To be precise, from the
supersymmetry transformations (\ref{SUSY}) for our ansatz (\ref{ads3s2}) 
one finds \cite{CD5}
\be
D=\frac{12}{l_S^2},\;\;\;\;\;\;\;p^I=-\frac{8}{3}VX^I,\;\;\;\;\;\;V=-\frac{3}{8}l_A,\;\;\;\;\;\;\;
l_A=2l_S.
\label{CON}
\ee 
On the other hand the equation of motion of auxiliary field $D$ is
\be
\frac{1}{6}C_{IJK}X^IX^JX^K+\frac{c_{2I}}{72}\left(DX^I+\frac{Vp^I}{l_S^4}\right)=1,
\ee
which, by making use of (\ref{CON}), can be recast to the following form
\be
\frac{1}{6}C_{IJK}X^IX^JX^K+\frac{1}{12l_A^2}c_{2I}X^I=1.
\ee
Setting $c_{2I}=0$, it reduces to $\N=1$ where one can define 
very special geometry underlines the theory at leading order. 
Therefore we would like to interpret this expression as a generalization of
$\N=1$ when the $R^2$ correction is also taken into account. This is analogous to the one loop
correction to the prepotential of four dimensional ${\cal N}=2$ supergravity which it is given by
$F=\frac{1}{6}\frac{C_{IJK}X^IX^JX^K}{X^0}+\Lambda^2\;\frac{c_{2I}X^I}{X^0}$.

 It is then natural to define the dual coordinates $X_I$ as 
\be
X_I=\frac{1}{6}C_{IJK}X^JX^K+\frac{1}{12l_A^2}c_{2I},
\ee
such that $X_IX^I=1$. From the supersymmetry conditions (\ref{CON}) one has $X^I=\frac{p^I}{l_A}$. Plugging
this into the above expressions we get $l_A^3={\frac{1}{6}C_{IJK}p^Ip^Jp^K+\frac{1}{12}c_{2I}p^I}$ and therefore
\be
X^I=\frac{p^I}{(\frac{1}{6}C_{IJK}p^Ip^Ip^K+\frac{1}{12}c_{2I}p^I)^{1/3}},\;\;\;\;\;\;
X_I=\frac{\frac{1}{6}C_{IJK}p^Ip^K+\frac{1}{12}c_{2I}}{(\frac{1}{6}C_{IJK}p^Ip^Ip^K+\frac{1}{12}c_{2I}p^I)^{2/3}}.
\ee
We will give more comments on this structure in the next section.

\section{Very special geometry and black holes at $R^2$ level}

In the previous section we have considered the black string solution in the presence of higher
derivative terms. We have observed that adding higher derivative terms will change the feature of
the very special geometry in such a way that the leading order constraint $\N=1$ is not satisfied any more.
Therefore the standard method of very special geometry is not applicable beyond tree level 
and we will have to solve the equations of motion coming from c-extremization directly.

In this section we would like to generalize the notion of very special geometry when
higher order corrections are also taken into account. It is possible to do that due to 
new progresses which have recently been made in computing the higher order correction using the fully 
supersymmetrized higher derivative terms. We note, however, that the notion of {\it generalized}
very special geometry depends on the explicit solution we are considering. To be
specified we will consider a black string solution in five dimensions  in the presence of
supersymmetric higher derivative terms and we define the generalized very special
geometry for this case. Then we will be able to reproduce all results of the previous section
in this framework. We will give a comment on how to generalize it for an arbitrary ansatz later
in the discussions section. 

Since the leading order constraint $\N=1$ defining the very special geometry is, indeed,
the equation of motion of the auxiliary field $D$, we will also define the
constraint for the generalized very special geometry by making use of the equation of motion of field $D$ 
in the presence of higher derivative terms. Using the black string ansatz of the previous section and taking
into account the supersymmetry constraints (\ref{CON}) the corresponding equation  
can be recast to the following form
\be
\frac{1}{6}C_{IJK}X^IX^JX^K+\frac{1}{12l_A^2}c_{2I}X^I=1.
\ee  
Now the main point to define the generalized very special geometry is as follows. When
we are considering the supersymmetrized action, in the ansatz we are choosing there is an auxiliary 
two-form field $v_{\mu\nu}$ which could be treated as an additional gauge field in the
theory with charge $p^0$. Accordingly, we could introduce new scalar field $X^0$ such that in the near 
horizon geometry one may set  $X^0=\frac{p^0}{l_A}$. Using this notation the above constraint may be written as
\be
\frac{1}{6}C_{IJK}X^IX^JX^K+\frac{1}{12}c_{2I}(X^0)^2X^I|_{p^0=1}=1.
\label{IJK}
\ee  
One may also define $F^0_{\theta\phi}=\frac{p^0}{2}\sin\theta$ such that 
$v_{\theta\phi}=-\frac{3}{4l_A}F^0_{\theta\phi}$. Obviously working with this notation all expressions 
reduce to those in section 3 for $p^0=1$, though it is not a solution for $p^0\neq 1$. Nevertheless 
we will work with $p^0\neq 1$ with the understanding that the solution is obtained
by setting $p^0=1$. It is worth noting that in general the auxiliary field 
$v_{\mu\nu}$ cannot be treated as a gauge field, though it can be seen that
for the models we are going to study it may be considered as a gauge field. 
Now we shall demonstrate how the solutions of previous section can 
be reproduced in this framework.
  
To start, let us first introduce new indexes $A,B,\cdots$ such that they take their values over 0 
and $I,J,\cdots$ by which the equation (\ref{IJK}) can be recast to the following form
\be
\frac{1}{6}C_{ABC}X^AX^BX^C=1
\label{ABC}
\ee
where $C_{ABC}=C_{IJK}$ for $A,B,C=I,J,K$ and $C_{00I}=\frac{c_{2I}}{6}$ and the other
components are zero. Thus, it is natural to define $X_A$ and the metric $C_{AB}$ as follows
\be
X_A=\frac{1}{6}C_{ABC}X^BX^C,\;\;\;\;\;\;\;\;\;\;\;\;\;\;\;C_{AB}=\frac{1}{6}C_{ABC}X^C.
\ee
More explicitly one has
\be
X_I=\frac{1}{6}C_{IJK}X^JX^K+\frac{c_{2I}}{36}(X^0)^2,\;\;\;\;\;\;\;\;\;\;\;\;X_0=\frac{c_{2I}X^I}{18}X^0,
\ee
and
\be
C_{IJ}=\frac{1}{6}C_{IJK}X^K,\;\;\;\;\;\;C_{I0}=\frac{c_{2I}}{36}X^0,\;\;\;\;\;\;C_{00}=\frac{c_{2I}X^I}{36}.
\ee
It is easy to verify that
\be
X_AX^A=1,\;\;\;\;\;\;\;\;X_A=C_{AB}X^B,\;\;\;\;\;\;\;\;C_{AB}X^AX^B=1.
\label{XX}
\ee
Following the standard notion of very special geometry, 
 the magnetic central charge may be defined as 
$Z_m=X_Ap^A$ and the near horizon parameters are fixed by extremizing it, {\em i.e.}
$\partial_iZ_m=\partial_iX_Ap^A=0$. Using (\ref{XX}) a solution would be $X^A=p^A/Z_m$.
Plugging this into (\ref{ABC}) we find
\be
Z_m=\left(\frac{1}{6}C_{ABC}p^Ap^Bp^C\right)^{1/3}=\left(\frac{1}{6}C_{IJK}p^Ip^Jp^K+\frac{(p^0)^2}{12}c_{2I}p^I
\right)^{1/3}.
\ee
To get our results in the previous section one needs to set $p^0=1$ at the end of the computations.

It is almost obvious, but  worth to mention, that having had the analogous to very 
special geometry does not mean that the action can also be given just by (\ref{5acts}) 
with the generalized constraint (\ref{ABC}). Therefore we would like to treat this
construction as an {\it auxiliary} method to study ${\cal N}=2$ supergravity
solutions in the presence of higher order corrections.

For example in the rest of this section we would like to show how the generalized very special geometry can be used to extend
the consideration of \cite{Guica:2007gm} while higher derivative terms are also taken into account.
To do this, consider the black string solution obtained in section \ref{3}. At $R^2$ level the solution 
can be written as follows
\bea
ds^2&=&\frac{l_A^2}{4}(dx^2-2rdxdr+\frac{dr^2}{r^2}+d\theta^2+\sin^2\theta\;d\phi^2),\;\;\;\;\;X^I=\frac{p^I}{l_A},\cr
A_\theta^I&=&-\frac{p^I}{2}\cos\theta\;d\phi,\;\;\;\;\;\;\;A^0_\theta=-\frac{1}{2}\cos\theta\;d\phi,\;\;\;\;\;
{\rm with}\;\;\;\;
v=-\frac{3}{4}l_AF^0_{\theta\phi},
\eea
where $l_A=\left(\frac{1}{6}C_{IJK}p^Ip^Jp^K+\frac{1}{12}c_{2I}p^I\right)^{1/3}$.  In writing the metric 
we have used the fact that $AdS_3$ can be written as $S^1$ fibered over $AdS_2$. Now we would like to embed
this solution into the framework of the generalized very special geometry. In order to do that, we start 
with the following solution with the understanding of $p^0=1$,
\bea
ds^2&=&\frac{Z_m^2}{4}(dx^2-2rdxdr+\frac{dr^2}{r^2}+d\theta^2+\sin^2\theta\;d\phi^2),\;\;\;\;\;\;
X^A=\frac{p^A}{Z_m},\cr
A_\theta^A&=&-\frac{p^A}{2}\cos\theta\;d\phi,\;\;\;\;\;\;\;Z_m^3=\frac{1}{6}C_{ABC}p^Ap^Bp^C.
\label{5d}
\eea
Following \cite{Guica:2007gm} we will consider the total space of $U(1)$ bundle over  (\ref{5d}) to
define a six dimensional manifold with the metric
\be
ds^2_6=\frac{Z_m^2}{4}(dx^2-2rdxdt+\frac{dr^2}{r^2}+\sigma_1^2+\sigma_2^2)+(2X_AX_B-C_{AB})(dy^A+A^A)(dy^B+A^B).
\label{6dads3}
\ee
Here $\sigma_i$ are right invariant one-forms such that $\sigma_1^2+\sigma_2^2=d\theta+\sin^2\theta\;d\phi^2$.
Let us define new coordinates $z,\psi$ through the following expressions
\be
y^A=z^A+(\sin B-1)X^A X_Bz^B-\frac{1}{2}p^A\psi,\;\;\;\;\;\;\;\;x=\frac{2\cos B}{Z_m}X_Az^A
\ee
where $\sin B$ is a constant which its physical meaning will become clear later. We define 
the coordinates such that the new coordinates have the following identification 
\be
\psi\sim\psi+4\pi m,\;\;\;\;\;\;\;\;\;z^A\sim z^A+2\pi n^A,
\ee
where $m$ and $n^A$ are integers. Accordingly one can read the identifications of $y$ and $x$.
Using the new coordinate $\psi$ one can define the right invariant one-forms by
\bea
&&\sigma_1=-\sin\psi d\theta+\cos\psi\sin\theta d\phi,\cr &&
\sigma_2=\cos\psi d\theta+\sin\psi\sin\theta d\phi,\cr 
&&\sigma_3=d\psi+\cos\theta d\phi.
\eea
In terms of the new coordinates the six dimensional metric (\ref{6dads3}) reads
\bea
ds^2&=&\frac{Z^2}{4}[-(\cos B\; rdt+\sin B \sigma_3)^2+\frac{dr^2}{r^2}+\sigma_1^2+\sigma_2^2+\sigma_3^2]\cr
 &&\cr &+&(2X_AX_B-C_{AB})(dz^A+{\tilde A}^A)(dz^B+{\tilde A}^B),
\eea
where $\tilde{A}^A=-\frac{p^A}{2}(\cos B\; rdt+\sin B \sigma_3)$. The obtained six dimensional manifold can be
treated as the total space of a $U(1)$ bundle over BMPV black hole at $R^2$ level. Therefore we can reduce
to five dimensions to get BMPV black hole where higher derivative corrections are also taken into account.
The resulting five dimensional black hole solution is
\bea
ds^2&=&\frac{l_A^2}{4}[-(\cos B\; rdt+\sin B \sigma_3)^2+\frac{dr^2}{r^2}+\sigma_1^2+\sigma_2^2
+\sigma_3^2],\;\;\;\;X^I=\frac{p^I}{l_A},\cr
&&\cr \tilde{A}^I&=&-\frac{p^I}{2}(\cos B \;rdt+\sin B \sigma_3),\;\;\;\;\;
\tilde{A}^0=-\frac{1}{2}(\cos B\; rdt+\sin B \sigma_3),
\eea
and the auxiliary field is given by $v=-\frac{3}{4}l_A {\tilde F}^0$. 
From this solution we can
identify $\sin B$ as the angular momentum, $J$, of the BMPV solution through the relation 
$\sin B=\frac{J}{Z^3}$. As a result we have
demonstrated that the total bundle space of near horizon wrapped M2's and wrapped M5's are 
equivalent up to $R^2$ level, generalizing the tree level results of \cite{Guica:2007gm}.
In fact one may go further to show that both of them can be obtained from quotients
of $AdS_3\times S^3$ with a flat $U(1)^{N-1}$ bundle, much similar to tree level
considered in \cite{Guica:2007gm}. It is then possible to use this connection to increase
our understanding of microstate counting of 5D supersymmetric rotating black hole 
arising from wrapped M2-branes in  Calabi-Yau compactification of M-theory. 
We hope to come back to this point in our future publication.

\section{Black hole solution}

In this section we shall consider a five dimensional extremal black 
hole\footnote{Explicit solutions of ${\cal N}=2$ 5D supersymmetric black holes 
in the context of very special geometry have been obtained for example in \cite{Chamseddine:1999qs}.} whose near
geometry is $AdS_2\times S^3$. When we are dealing with an extremal black hole with $AdS_2$ near 
horizon geometry, it is more appropriate to work with entropy function formalism \cite{Sen:2005wa}.
In fact this approach has been used to study five dimensional extremal black hole in Heterotic string
theory in presence of higher derivative terms given by Gauss-Bonnet action 
\cite{Sen:2005iz}.\footnote{Higher order corrections to 5D BH have also been studied
in \cite{{Nojiri:2001ae},{Ish},{Guica:2005ig}}.}
It is shown that this {\it bosonic} term was enough to correctly reproduce the microscopic
entropy coming from microstate counting in string theory. It is the aim of this section
to study the five dimensional extremal black hole in the presence of higher derivative terms which
come from supersymmetrized action. We will also compare the result with the case where only
bosonic Gauss-Bonnet term is present. In fact it is analogous to the consideration of 
\cite{{Sahoo:2006rp},{Alishahiha:2006jd}} where four dimensional extremal black hole has been
studied in the presence of supersymmetrized action using entropy function formalism.  

Let us start with the following ansatz for near horizon geometry
\be
ds=l_A^2ds_{ADS_2}^2+l_S^2ds_{S^3}^2,\;\;\;\;\;X^I=cont.\;\;\;\;\;\;F^I_{rt}=e^I,
\;\;\;\;\;\;v_{rt}=V.
\label{ads2s3}
\ee
Then the entropy function  is given by ${\cal E}=2\pi(e^Iq_I-f_0+f_1)$ where $f_0$ is the
leading order contribution coming from quadratic part of the action which  is 
\be
f_0=\frac{1}{2}l_A^2l_S^3\bigg{[}\frac{\N-1}{2}D+\frac{\N+3}{2}\left(\frac{3}{l_S^2}-\frac{1}{l_A^2}\right)
-\frac{2(3\N+1)}{l_A^4}V^2-\frac{4\N_Ie^I}{l_A^4}V-\frac{\N_{IJ}e^Ie^J}{2l_A^4}\bigg{]},
\ee
and the higher order contribution, $f_1$, comes from the four-derivative terms that for our 
ansatz, is
\bea
f_1&=&\frac{c_{2I}}{48}l_A^2l_S^3
\bigg{[}\frac{X^I}{4}\left(\frac{1}{l_S^2}-\frac{1}{l_A^2}\right)^2+\frac{4V^4}{l_A^8}X^I
+\frac{4V^3}{3l_A^8}e^I-\frac{DV}{3l_A^4}e^I+\frac{D^2}{12}X^I\cr &&\cr &&\cr
&&\;\;\;\;\;\;\;\;\;\;\;\;\;\;-\frac{2V^2X^I}{3l_A^4}\left(\frac{3}{l_S^2}+\frac{5}{l_A^2}\right)-\frac{Ve^I}{l_A^4}
\left(\frac{1}{l_A^2}-\frac{1}{l_S^2}\right)\bigg{]}.
\eea

At leading order where only $f_0$ contributes, one may integrate out the
the auxiliary fields by extremizing the entropy function with respect to them, arriving at
\be
f_0=\frac{1}{2}l_A^2l_S^3\left[\frac{6}{l_S^2}-\frac{2}{l_A^2}+\frac{1}{2l_A^4}(\N_I\N_J-\N_{IJ})e^Ie^J\right]
=\frac{1}{2}l_A^2l_S^3\left(\frac{6}{l_S^2}-\frac{2}{l_A^2}+\frac{G_{IJ}e^Ie^J}{l_A^4}\right).
\ee
It is easy to extremize the entropy function with respect to the
parameters to find $l_A,l_S,X^I$ and $e^I$. In particular for STU model where
$C_{123}=1$ and the other components are zero one gets
\be
l_S=2l_A=(q_1q_2q_3)^{1/6},\;\;\;
X^I=\frac{(q_1q_2q_3)^{1/3}}{q_I},
\;\;\;e^I=\frac{1}{2}
\frac{(q_1q_2q_3)^{1/2}}{q_I},
\ee
and the corresponding black hole entropy is $S=2\pi\sqrt{q_1q_2q_3}$.

In general it is difficult to do the same while the higher order corrections are also taking into account.
Therefore to proceed, the same as in the previous, we will use the supersymmetry transformations to 
simplify as much as we can and then using the equation of motion of field $D$, we find a supersymmetric 
solution. For our ansatz (\ref{ads2s3}) the supersymmetry conditions (\ref{SUSY}) lead to
\be
D=-\frac{3}{l_A^2},\;\;\;\;\;\;\;e^I=-\frac{4}{3}VX^I,\;\;\;\;\;\;V=-\frac{3}{4}l_A,
\;\;\;\;\;\;l_S=2l_A.
\label{con}
\ee 
Using these relations we may set $X^I=\frac{e^I}{l_A}$ and defining $E=\frac{1}{6}C_{IJK}e^Ie^Je^K$ one has 
\be
\N=\frac{1}{l_A^3}E,\;\;\;\;\;\;\N_Ie^I=\frac{3}{l_A^2}E,\;\;\;\;\;\;\N_{IJ}e^Ie^J=\frac{6}{l_A}E.
\ee
In this notation the equation of motion for auxiliary field $D$ reads
\be
E-l_A^3+\frac{l_A^3}{12}c_{2I}\left(\frac{DX^I}{6}-\frac{Ve^I}{3l_A^4}\right)=0
\ee
so that $l_A=\frac{1}{2}(8E-\frac{c_{2I}e^I}{6})^{1/3}$. By making use of the expressions for the
parameters $D,V,X^I$ and $l_S$ given above, the entropy function gets the following simple form
\be
{\cal E}=2\pi(q_Ie^I-4E+\frac{1}{8}c_{2I}e^I)=2\pi\left[(q_I+\frac{1}{8}c_{2I})e^I-\frac{2}{3}C_{IJK}e^Ie^Je^K\right]
\label{EN}
\ee
Extremizing the entropy function with respect to $e^I$ we get
\be
2C_{IJK}e^Je^K=q_I+\frac{1}{8}c_{2I}
\ee
which in principle can be solved to find $e^I$ in terms of the charges, $c_{2I}$ and
parameters $C_{IJK}$'s. The entropy is also given by
 $S=\frac{4\pi}{3}q^+_Ie^I$ \footnote{We use a notion in 
which $q^+_I=q_I+\frac{1}{8}c_{2I}$.}.
 In particular for STU model we get $
e^I={(q^+_1q^+_2q^+_3)^{1/2}}/2{q_I^+}$.
 Plugging this into the expressions of $l_A$ and $X^I$, we obtain
\be
2l_A=(q^+_1q^+_2q^+_3)^{1/6}(1-\frac{c_{2I}}{12}\frac{1}{q_I^+})^{1/3}
,\;\;\;\;\;\;\;\;\;
X^I=\frac{(q_1^+q_2^+q_3^+)^{1/3}}{q_I^+}(1-\frac{c_{2I}}{12}\frac{1}{q_I^+})^{-1/3}.
\ee
Finally the entropy is found to be
\be
S=2\pi\sqrt{q^+_1q^+_2q^+_3}.
\ee
It is very interesting in the sense that the entropy of the black hole at $R^2$ level has the 
same form as the tree level except that the charges $q_I$'s are replaced by shifted charges $q^+_I$'s.

This result can be used to see how the higher order corrections stretch horizon.
For example if we start by a classical solution with $q_1=0$ we get vanishing 
horizon (small black hole) and therefore the entropy is zero. 
But adding the $R^2$ terms we get a smooth solution with a non-zero entropy give by
\be
S=\pi\sqrt{\frac{c_{21}}{2}q^+_2q^+_3}.
\ee

This procedure could also be used to understand, upon dimensional reduction to four dimensions,  the single-charge small black hole studied in \cite{Sinha:2006yy}.

We note that in the above consideration we have used the supersymmetry transformations to simplify the
computations and therefore the solution would be supersymmetric. Thus it does not exclude other solutions. 
In fact we would expect to have another solution corresponding
to non-BPS solution. Actually one could start from a more general constraint than (\ref{con}) as follows
\be
D=\frac{\alpha}{l_A^2},\;\;\;\;\;\;\;X^I=\beta\frac{e^I}{l_A},\;\;\;\;\;\;V=\gamma l_A,
\;\;\;\;\;\;l_S=2l_A
\ee 
and solve the equations for parameters $(\alpha,\beta,\gamma)$. Doing so, for fixed $l_A$ given above,
we find two solutions: $(-3,1,-3/4)$ and $(3,-1,-3/4)$. The first one is the solution we have studied, but
the second one which is not supersymmetric leads to the following entropy function
\be
{\cal E}=2\pi\left[(q_I-\frac{3}{8}c_{2I})e^I-\frac{2}{3}C_{IJK}e^Ie^Je^K\right].
\ee
It is straightforward to extremize the entropy function with respect to $e^I$'s to get the 
parameters in terms of $q_I$'s, though we won't do that here.

It is also instructive to compare the results with the case where the higher order corrections are given
in terms of the Gauss-Bonnet action. It is known that this term cannot be supersymmetrized, but it is still interesting to see what would be  the corresponding corrections. In our notation the Gauss-Bonnet term is given by
\be
{\cal L}_{GB}=\frac{c_{2I}X^I}{2^8\cdot 3\pi^2}\left(R^{abcd}R_{abcd}-4R^{ab}R_{ab}+R^2\right).
\ee
It can be shown that with the specific coefficient we have chosen for the Gauss-Bonnet action, the corresponding 
entropy function, fixing the auxiliary fields as (\ref{con}) for tree level action, is the same as
(\ref{EN}) and therefore we get the same results as those in supersymmetrized action.

\section{Discussions}
In this paper we have studied black string and black hole solutions in the presence of higher 
derivative terms. In the black string case, where the near horizon geometry is
$AdS_3\times S^2$, a proper method is c-extremization, whereas in the black hole solution
with $AdS_2\times S^3$ near horizon geometry the better method is given by the entropy function formalism.
In both cases we have seen that adding higher derivative terms will change the feature of
the very special geometry in such a way that the leading order constraint $\N=1$ is not satisfied any more.
Therefore the standard method of very special geometry is not applicable at this level.

We have shown how to generalize the notion of very special geometry in the presence of 
higher derivative terms. We have observed that the generalized very special geometry 
depends on the solution we are considering. In particular we have explicitly studied
the generalization for the magnetic black string solution where we have reproduced
all results obtained from another method, {\it e.g.} c-extremization. 

For an arbitrary solution, the generalized constraint which defines the very special geometry can be obtained from
equation of motion of the auxiliary filed $D$. In general we have
\be
\frac{1}{6}C_{IJK}X^IX^JX^K+\frac{ c_{2I}}{72}(X^ID+F^{I\mu\nu}v_{\mu\nu})=1.
\ee
If we are interested in a supersymmetric solution, we can also use the supersymmetry transformations
to further simplify the constraint 
\be
\frac{1}{6}C_{IJK}X^IX^JX^K+
\frac{ c_{2I}X^I}{54}\left(\frac{3}{2}\epsilon_{abcde}\gamma^av^{bc}v^{de}-(\gamma\cdot v)^2
-v^2\right)=1.
\ee

An interesting application of this construction is to generalize the consideration of \cite{Guica:2007gm}.
In particular we have shown that the total bundle space of near horizon wrapped M2's and wrapped M5's are 
equivalent up to $R^2$ level, generalizing the tree level results of \cite{Guica:2007gm}.
Actually we could also show that both of them can be obtained from quotients
of $AdS_3\times S^3$ with a flat $U(1)^{N-1}$ bundle, much similar to tree level
considered in \cite{Guica:2007gm}. Although we have not pushed this observation any further, one
might suspect that using this connection could increase
our knowledge of microstate counting of 5D supersymmetric rotating black hole arising from
wrapped M2-branes in  Calabi-Yau compactification of M-theory. 

We have also studied higher order corrections to extremal black holes in five 
dimensions in which the higher order corrections come  from supersymmetric 
completion of $R^2$ term. In this paper we have explicitly written the corrected 
solution of 5D black holes for a specific model, namely, STU model. 
To extend the solution for a general model we will have to
solve a set of equations
\be
C_{IJK}x^Jx^K=a_I,\;\;\;\;\;\;\;I=1,\cdots, N,
\label{yyy}
\ee
for given $C_{IJK}$ and $a_I$. In general it is difficult to solve this equation and
it may not even have a unique solution. Nevertheless for a particular model 
this can be solved explicitly (for example see \cite{Shmakova:1996nz}).
An interesting observation we have made in this paper is that the solution of the
above equation can be used for both tree level case and the case when the $R^2$ 
corrections are added.
The only difference is that in the presence of $R^2$ terms one just needs to replace 
$a_I$ by another constant which is related to it by a constant shift.

In fact as far as the entropy is concerned we have seen that the corrected entropy can simply be obtained by replacing the electric charges $q_I$'s with the shifted charges $q_{I}^+$'s defined in footnote 2. In particular this result shows how one can get a smooth solution with non-zero entropy out of a small black hole  which in leading order
supergravity is singular with vanishing entropy.

Finally as an aside let us define $C^{IJK}$ in terms of $C_{IJK}$ by 
$C^{IJK}C_{KLM}=\delta^{I}_L\delta^J_M+\delta^{I}_M\delta^J_L$. Although it is
not clear whether this equation can be solved in general or even if it has a unique solution, it 
can be used to write a solution for equation (\ref{yyy}). For example it terms
of this parameter the corresponding black hole entropy could be written as \cite{Larsen:2006xm}
\be
S=2\pi\sqrt{\frac{1}{6}C^{IJK}q_Iq_Jq_K}.
\ee
In the presence of $R^2$ correction the entropy has the same form, except that
one needs to replace $q_I$ by $q^+_I$.

{\bf Note added}: While we were in the final stage of the project we have received 
the paper \cite{new} where the five dimensional black hole in the presence 
of higher derivative terms given by (\ref{act5dR2}) has also been studied.

{\bf Acknowledgments}

I would like to thank Hajar Ebrahim for discussions on the related topic and also comments
on the draft of the article.

\end{document}